\documentclass[aps,twocolumn,showpacs]{revtex4}
\usepackage{graphics,epsfig,epsf,psfrag}
\usepackage{bm}
\usepackage{color,xcolor}
\usepackage{amsmath}
\usepackage{amssymb}
\usepackage{amsfonts}
\usepackage{epstopdf}
\usepackage{isotope}
\usepackage{dcolumn}
\def\vec#1{\boldsymbol{#1}}
\begin{document}

\title{The lightest neutral hypernuclei with strangeness $-1$ and $-2$}

\author{Jean-Marc Richard$^1$\footnote{j-m.richard@ipnl.in2p3.fr}, Qian Wang$^2$\footnote{q.wang@fz-juelich.de}, and Qiang Zhao$^3$\footnote{zhaoq@ihep.ac.cn}}

\affiliation{
       $^1$Universit\'e de Lyon, Institut de Physique Nucl\'eaire, UCBL--IN2P3-CNRS,
4, rue Enrico Fermi, Villeurbanne, France\\
        $^2$Institut f\"{u}r Kernphysik, Institute for Advanced
Simulation, and J\"ulich Center for Hadron
Physics, D-52425 J\"{u}lich, Germany\\
        $^3$\it Institute of High Energy Physics and Theoretical Physics Center for Science Facilities,\\
        Chinese Academy of Sciences, Beijing 100049, China
          }

\begin{abstract}
Our current knowledge of the baryon--baryon interaction suggests that the dineutron $(n,n)$ and its strange analogue $(n,\Lambda)$ are unstable. In contrast, the situation is more favorable for  the strange three-body system $(n,n,\Lambda)$, and even better for the four-body system  $T\equiv (n,n,\Lambda,\Lambda)$ with strangeness $-2$, which is likely to be stable under spontaneous dissociation.
The recent models of the hyperon-nucleon and
hyperon-hyperon interactions suggest that the stability of the
$(n,n,\Lambda)$ and $T$ is possible within the uncertainties
of our knowledge of the baryon-baryon interactions.
This new nucleus $T$ could be produced and identified in central deuteron--deuteron collisions via reaction $d+d\to T+K^++K^+$, and the tetrabaryon $T$ could play an important role in catalyzing the formation of a strange core in neutron stars.
\end{abstract}

\maketitle

\section{Introduction}
It is generally accepted that the dineutron
$\isotope[2]{n}=(n,n)$ and its strange analogue $\isotope[2][\Lambda]{n}=(n,\Lambda)$, involving the lightest hyperon, are unbound. The situation is more complicated for some of  the 3-body or 4-body systems made of nucleons and hyperons, and needs to be clarified. The tetraneutron, $\isotope[4]{n}=(n,n,n,n)$, is naively suggested by the stability  of \isotope[8]{He} isotope, but has received only controversial experimental indications~\cite{Marques:2001wh}. Its stability cannot be established using realistic neutron-neutron potentials \cite{Tang:1965zza,Bertulani:2002px,Pieper:2003dc,Timofeyuk:2003ya,Lazauskas:2005ig}, nor even with potentials made artificially deeper to produce a dineutron  \isotope[2]{n}, due to the Pauli principle.  The question is thus addressed of the stability of a modified tetraneutron with Bose statistics, namely $T=\isotope[4][\Lambda\Lambda]{n}=(n,n,\Lambda,\Lambda)$ with spin-singlet wavefunctions for both the $nn$ and the $\Lambda\Lambda$ pairs
endorsing the antisymmetrisation.

The physics of hypernuclei is progressing dramatically in both experiment and theory, for which recent reviews can be found in \cite{snp12,Gal:2011tb}.
For baryon number $A=2$, there is no evidence for a stable $(n,\Lambda)$ nor $(p,\Lambda)$ bound state, except for a mass peak of about 2.06\,GeV seen in $d+\pi^-$ \cite{Rappold:2013jta}.
On the other hand, for $A=3$, $\isotope[3][\Lambda]{H}=(n,p,\Lambda)$ is bound with an energy of  $E_3=-2.45$ MeV, i.e., just below the threshold for separation in a deuteron plus an isolated hyperon, at $E_2=-2.20\,$ MeV. There is only one claim for $\isotope[3][\Lambda]{n}=(n,n,\Lambda)$, at about 3\,GeV~\cite{Rappold:2013jta}.

The sector of strangeness $S=-2$, or ``double $\Lambda$'' hypernuclei has also been investigated, in particular, after the ``Nagara'' event \cite{Takahashi:2001nm}, which is the very accurate recent observation of~\isotope[6][\Lambda\Lambda]{He}, and a measurement of its binding energy, which sets a limit on the $\Lambda\Lambda$ effective attraction in that system. This Nagara event motivated to question whether lighter $S=-2$ systems exist or not~\cite{Garcilazo:2012qv,Gal:2013bw,Garcilazo:2013zwa}.

In 2002, Filikhin and Gal~\cite{Filikhin:2002wp} studied the $\isotope[4][\Lambda\Lambda]{H}$ system, and found it unbound within the models they adopted. However, their calculation was revisited by Nemura, Akaishi and Myint \cite{Nemura:2002hv}, who used a more sophisticated method for solving the four-body problem, and found a small amount of binding, below the dissociation threshold into $\isotope[3][\Lambda]{H}+\Lambda$.
This illustrates once more how the four-body problem is delicate in the regime of weak binding.
In Refs.~\cite{Filikhin:2002wp,Nemura:2002hv,Garcilazo:2012qv}, it is stressed that
the free $\Lambda\Lambda$ interaction receives a significant contribution from  the $\Lambda\Lambda\leftrightarrow N\Xi\leftrightarrow \Lambda\Lambda$ coupling, which is mediated by the exchange of kaons, and  is suppressed in a dense nucleus due to the antisymmetrisation  between the nucleons in the core and the nucleon in $N\Xi$. This Pauli suppression was successfully invoked to explain the relatively weak binding of $\isotope[6][\Lambda\Lambda]{He}$. However, the Pauli suppression requires a $(n\Lambda\Lambda)$ correlation inside the $\isotope[6][\Lambda\Lambda]{He}$ system, and thus it does not operate  in the limit of very weak binding.

The aim of this work is to investigate the stability of the \isotope[4][\Lambda\Lambda]{n} system, to discuss its production mechanism and its possible role in astrophysics. As follows, we will demonstrate in Sec. ~\ref{binding-cond} that \isotope[4][\Lambda\Lambda]{n} is likely a ``Borromean'' system if it does exist. In Sec.~\ref{prod-mech} we propose and discuss its production mechanism in deuteron-deuteron scatterings. A summary is then given in Sec.~\ref{summ}.

\section{Analysis of binding conditions for 3 and 4-body systems with strangeness $-1$ and $-2$}\label{binding-cond}

We shall start by some reminders of binding systems whose subsystems are unbound in this Section. In his study of the range of nuclear forces, Thomas \cite{Thomas:1935zz} discovered that the ratio of 3-body to 2-body bound-state energies, $E_3/E_2$, becomes very large if the range of the interaction decreases. Equivalently, for a given (short) range, $E_3/E_2\to\infty$ if the coupling $g$ approaches (from above) the minimal value $g_2$ required to bind two particles. Here, the potential energy is written as $g\,\sum v(r_{ij})$, where $v$ is attractive or contains attractive parts, and $r_{ij}$ denotes the inter-particle separation. Implicit in this ``Thomas collapse'' is that the minimal coupling $g_3$ to bind three particles is smaller than $g_2$, so that for $g_3<g<g_2$ there is a ``Borromean'' 3-body bound state whose two-body subsystems are unbound.

Since the work by Thomas in 1935, Borromean binding has been further investigated. There are rigorous bounds on the allowed domain of coupling constants for binding systems whose subsystems are unbound~\cite{Richard:1994mc}, and studies on how wide is the Borromean window as a function of the shape of the potential~\cite{Moszkowski:2000ni} or of the low-energy parameters of the pair interactions \cite{PhysRevA.78.020501}.

\begin{figure}[!htbp]
\centerline{
 \hspace*{-10pt}
    \includegraphics[width=0.40\linewidth]{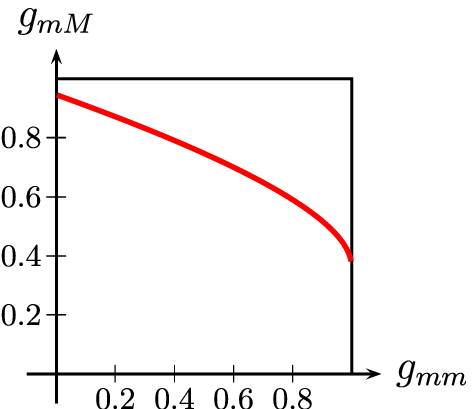}
    \hspace*{10pt}
    \includegraphics[width=0.49\linewidth]{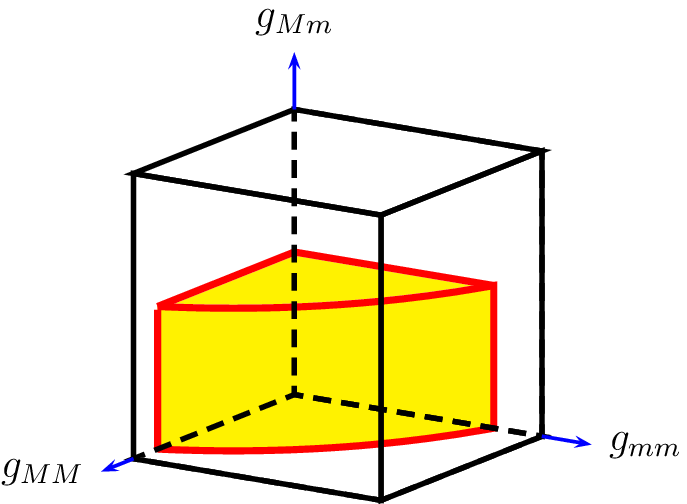}}
\vskip 4pt
\centerline{\hfill (a)\hfill\hfill (b)\hfill}
\vskip 4pt
\caption{(Color online) Domain in which binding is forbidden for a $(m,m,M)$ (a) or  $(m,m,M,M)$  (b) system. The scale is set such that the pair $(m_i,m_j)$ is bound for $g_{ij}>1$. This drawing corresponds to $M/m=1.2$.}
\label{fig:Mmm-MMmm}
\end{figure}

The allowed values of $g_3/g_2$ can  be bound below rigorously~\cite{Richard:1994mc}. The basic idea is to decompose the Hamiltonian $H$ into sub-Hamiltonians, say $H=H_1+H_2+\cdots$. Then $H$ hardly becomes negative if all $H_i$ are positive. There are several variants and refinements, depending on whether the center-of-mass motion is properly removed and enough flexibility is introduced to account for unequal masses and couplings. For the three-body problem, the best decomposition \cite{Basdevant:1992cm} leads for $(m,m,M)$ to the forbidden domains shown in Fig.~\ref{fig:Mmm-MMmm}\,(a).

As seen in Fig.~\ref{fig:Mmm-MMmm}\,(a), the forbidden domain extents to nearly 3/4 of the Borromean sector
$\{ g_{31}=g_{12}\le 1,\, g_{23}\le 1\}$. The reason is that as far as one deviates from the case of three identical particles, the various cluster decompositions of the wavefunction, such as $[(12)3]$ or $[(23) 1]$, would not overlap much and cannot interfere efficiently to build  a collective binding of these three particles.

The same reasoning, when applied to  the $(m,m,M,M)$ system that involves two masses and three couplings, gives
the forbidden domain~\cite{Richard:1994mc} shown in Fig.~\ref{fig:Mmm-MMmm}\,(b). The allowed fraction  is larger than in the 3-body case. This is an encouragement for us to perform the 4-body calculations, namely, to determine to which extent some points of the ``non-forbidden'' domain can lead to actual 4-body bound states. Another encouragement is the existence of  fully Borromean 4-body bound states in the case of screened Coulomb interaction \cite{PhysRevA.69.042504}.

The actual frontier of Borromean binding depends, indeed,  on the shape of the potential. An analysis  has been proposed in~\cite{Moszkowski:2000ni}. The rigorous lower bound tends to be saturated for potentials with an external repulsive barrier. On the other hand, the window of Borromean binding is more and more reduced for potentials with harder and harder inner core.

\begin{figure}[!htbp]
\centering
\includegraphics[width=0.8\linewidth]{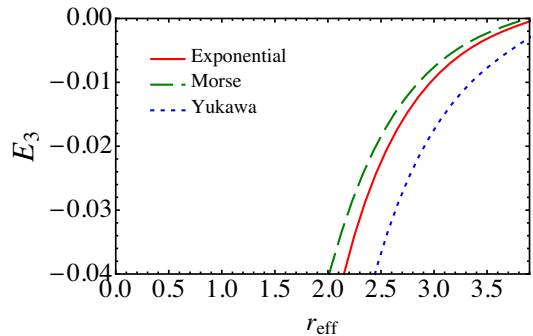}
\caption{(Color online) Three-body energy $E_3$ as a function of the 2-body effective range parameter $r_\text{eff}$, for the exponential, Yukawa and Morse (with $R=0.6$) potentials,  where $g$ and $\mu$  are linked to reproduced a 2-body scattering length $a=-10$.}
\label{fig:comp}
\end{figure}

In Fig.~\ref{fig:comp}, a slightly different analysis is proposed. For a mass $m$ such that
$m/\hbar^2=1$ to fix the scale, we compare the 3-body binding energy $E_3$ as a function of the effective range $r_\text{eff}$ for three different potentials, i.e.,
\begin{align}
& -g\,\exp(-\mu\,r)\qquad \text{(exponential)}~, \label{eq:exp}\\
&-g\,\exp(-\mu\,r)/r\quad\text{(Yukawa)}~, \label{eq:yuk}\\
& g \exp[-2\,\mu\, (r-R)]-2\,g\,\exp[-\mu\,(r-R)] \  \text{(Morse)}~, \label{eq:Morse}
\end{align}
where the 2-body scattering length is fixed to a large negative value $a=-10$, i.e., just below the threshold for 2-body binding. This gives $g$ as a function of $\mu$. For the Morse potential, we use $R=0.6$ for the illustration purpose.

The lessons of this exercise are:
\begin{itemize}

\item The algebraic energy is a sharply increasing function of the effective range. This means that models with a very large effective range cannot generate much Borromean binding~\cite{PhysRevA.78.020501}. Note that in the case of ordinary 3-body binding (not fully Borromean), the dependence upon the effective range might be different~\cite{Gibson:1977ch,Gal:2014efa}.

\item The amount of Borromean binding does not depend dramatically on the shape of the potential, as seen by comparing models whose short-range part is attractive (Yukawa), constant (exponential) or repulsive (Morse).  The behavior would be universal, if instead of $E_3$ vs.\ $r_\text{eff}$, one plots $E_3$ against the imaginary part of the pole position
of the 2-body problem.
However, since the pole position is not easily accessible in the models we use,  we keep the effective range as a tool for the discussion.

\end{itemize}

In short, it looks reasonable to use simple models to study Borromean binding, given that both the scattering length and effective range are properly reproduced.

For Fig.~\ref{fig:comp} and other 3-body problems in the regime of weak binding, an efficient variational method makes use of the distances $x=|\vec r_2-\vec r_3|$, \dots, and a basis of exponential functions
$\Psi=\sum_i \alpha_i\,\exp[-(a_i\,x+b_i\,y+c_i\,z)]$. For any given set of  non-linear parameters $a_i, \,b_i, \,c_i$, the variational energy and coefficients $\alpha_i$ are obtained from  a simple eigenvalue equation. As done sometimes for Gaussians~\cite{Hiyama:2003cu}, the non-linear parameters are assigned to belong to a single geometric series $a,\,a\,v, \,a\,v^2,\,\ldots$, and thus the minimisation of the energy is achieved by varying $a$ and $v$.
The results are cross-checked against the stochastic variational method of Suzuki and Varga~\cite{Suzuki:1998bn} based on Gaussian wavefunctions,  and the associated computer code \cite{Varga:1997xga}, which is also used for 4-body systems.

We first test a simple toy model, where the potential for each pair is taken as $-g_{ij}\,\exp(-\mu_0\,r)$, with a unique  range $\mu_0^{-1}$. The 2-body problem can be solved analytically to identify the couplings $g_{ij}$ providing a given 2-body energy or a given 2-body scattering length. Once the nucleon-nucleon sector is fixed, the $g_{N\Lambda}$ coupling of the nucleon-hyperon interaction, assumed to be spin-independent, has been fixed as to reproduce an energy of  about  $-2.45\,$MeV for \isotope[3][\Lambda]{H}.

For $\mu_0=1\,$GeV, i.e., a short-range toy model, one obtains unrealistically-low values for the effective-range parameters, e.g., about $3.2\,$GeV$^{-1}$ for $(np)$ and $3.6\,$GeV$^{-1}$ for $(nn)$, and the 4-body configurations of interest are found stable. In particular, $\isotope[4][\Lambda\Lambda]{H}$ is bound below the threshold for dissociation into $\isotope[3][\Lambda]{H}+\Lambda$, even for a coupling $g_{\Lambda\Lambda}$ as low as about a quarter of the value required to bind $(\Lambda\Lambda)$, and the isovector configurations $\isotope[4][\Lambda\Lambda]{H}$ and~$\isotope[4][\Lambda\Lambda]{n}$ are found stable against complete dissociation.

In contrast, for $\mu_0=0.2\,$GeV corresponding to a long-range interaction, the effective-range parameters are systematically over-estimated, and the 4-body systems $\isotope[4][\Lambda\Lambda]{H}$ with both isospsin $I=0$ and $I=1$, and $\isotope[4][\Lambda\Lambda]{n}$ become unstable, unless a large value is adopted for $g_{\Lambda\Lambda}$, implying a positive $\Lambda\Lambda$ scattering length and a bound $\Lambda\Lambda$.

The  analysis with toy models motivates the choice of a tractable potential which gives realistic effective-range parameters, i.e., which reproduces either the  scattering lengths $a_\text {sc}$ and effective-range $r_{\rm eff}$  of  the models ESC08a-ESC08c of the Nijmegen-RIKEN group \cite{snp12,Rijken:2010zzb,Rijken:2013wxa}, or the more recent values based on chiral effective field theory (CEFT)\cite{Polinder:2007mp,Haidenbauer:2013oca}.
More precisely, we adopted the  parameters given in Table~\ref{tab:tab1}.
\begin{table}[!htbc]
\caption{ Values (in fm) adopted for the scattering length and effective range parameter in the two models.}\label{tab:tab1}
\begin{ruledtabular}
\begin{tabular}{cdddd}
 & \multicolumn{2}{c}{ESC08} & \multicolumn{2}{c}{CEFT}\\[-2pt]
Pair & a &  r_\text{reff} &  a & r_\text{eff}\\
\hline\\[-5pt]
$nn$ &-16.51 & 2.85 &-18.9 & 2.75\\
$(n\Lambda)_{s=0}$ &-2.7 &2.97 & -2.9&2.65\\
$(n\Lambda)_{s=1}$ &-1.65& 3.63& -1.51 &2.64\\
$\Lambda\Lambda$ & -0.88 &4.34 &-1.54&0.31
\end{tabular}
\end{ruledtabular}
\end{table}


For the potential, we adopted the exponential shape \eqref{eq:exp}, and the Morse interaction \eqref{eq:Morse}, with a reasonable range $R=0.6\,$fm. The results are similar, with slightly more binding for the exponential parametrisation, as already observed in Fig.~\ref{fig:comp}.

The model reproduces, not surprisingly, the observed binding of the \isotope[2]{H}\ and \isotope[3][\Lambda]{H}\ systems. For the latter, both the spin $s=1/2$ and $s=3/2$ exist, as there is no much difference between the spin-triplet and the spin-averaged nucleon-hyperon interactions. The states \isotope[3][\Lambda]{H} with isospin $I=1$ and spin $s=1/2$ and  $\isotope[3][\Lambda]{n}$ are marginally unbound, and would become bound, for instance, if some masses are increased by about 10\%. Our results for $\isotope[3][\Lambda]{n}$ agree with the conclusions of recent studies \cite{Gal:2014efa,Hiyama:2014cua,Valcarce2014}. The state \isotope[4][\Lambda\Lambda]{H} with isospin $I=0$ is found weakly bound (about 3\,MeV) in the Nijmegen-RIKEN model, and slightly more (about 9\,MeV) in the CEFT one.  The state \isotope[4][\Lambda\Lambda]{H} with isospin $I=1$ and \isotope[4][\Lambda\Lambda]{n} miss binding by a very small amount with the Nijmegen-RIKEN parameters, but become bound by about 1\,MeV with the CEFT
parameters. It is interesting that the most recent approach to inter-hadronic forces slightly differs from the conventional approach.

At this point, however, one cannot firmly conclude before including 3-body forces. They probably have an attractive component, which is the analogue of the attraction brought by 3-body forces to few-nucleons systems.
Some spin-dependence of the 3-body forces cannot be excluded, this could be invoked to keep the spin $s=1/2$ state of \isotope[3][\Lambda]{H} bound and move the $s=3/2$ in the continuum.
The 3-body and $n$-body forces with $n>3$  probably contain a repulsive-component of shorter-range, which reflects that if several hyperons or several hyperons and nucleons overlap, they feel the effect of the Pauli exclusion of their constituent quarks. This repulsive component seems needed in large systems containing strangeness~\cite{Lonardoni:2013rm}. The calculations of hyperon-nucleon and hyperon-hyperon forces should be pushed to higher order within chiral effective theories and Nijmegen-RIKEN models, as the 3-body components will emerge automatically together with a refinement of the 2-body ones.

\section{Production mechanism for $\isotope[4][\Lambda\Lambda]{n}$ }\label{prod-mech}

We now propose and discuss a possible production mechanism for $T=\isotope[4][\Lambda\Lambda]{n}$, but omitting most  details.
The most ideal process to create  such a loosely-bound neutral system is via the deuteron-deuteron ($d-d$) scattering, i.e., $d+d\to K^+ + K^+ +T$.
This is an extremely clean process. By tagging the $K^+K^+$ events and examining the  missing mass spectrum recoiling against the $K^+K^+$ pairs, the signal for $T$, if $T$ does exist,  will accumulate at the mass of $T$ to form a narrow peak. We now analyze the reaction mechanism and make a numerical evaluation of the cross section.

\begin{figure}[!htbp]
\centering
\includegraphics[width=0.9\linewidth]{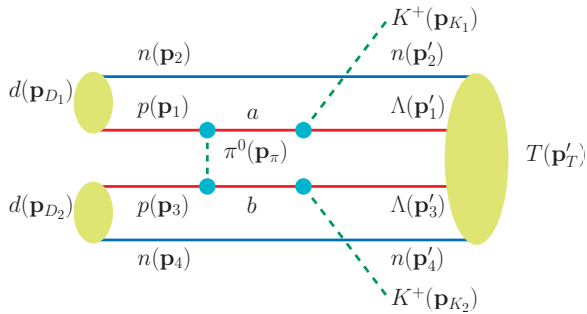}
\caption{(Color online) Mechanism for $T$ production in $dd$ collisions.}
\label{fig:reac}
\end{figure}

In Fig.~\ref{fig:reac} one of the typical transition processes is illustrated with the kinematic variables defined and others are implicated. In the $d-d$ collision, the two protons scatter and exchange mesons such as $\pi^0$, $\eta$, $\rho^0$ etc, and the $K^+K^+$ pair is created in association with the $\Lambda\Lambda$. The central collision will kinematically favour the formation of
$(n,n,\Lambda,\Lambda)$ in the final state. As the leading-order test of this formation mechanism, we assume that the relative internal momenta between the proton and neutron inside the incident deuterons is negligibly small compared to the relative momentum between these two initial deuterons. It means that we have neglected the Fermi motion inside the initial deuterons in the reaction. We have also neglected the final-state interactions among the final-state baryons. Thus the neutrons are treated as spectators and their contributions to the amplitude will be via the convolution of the final-baryon momentum distributions.

\begin{figure}[!htbp]
\centering
\includegraphics[width=0.8\linewidth]{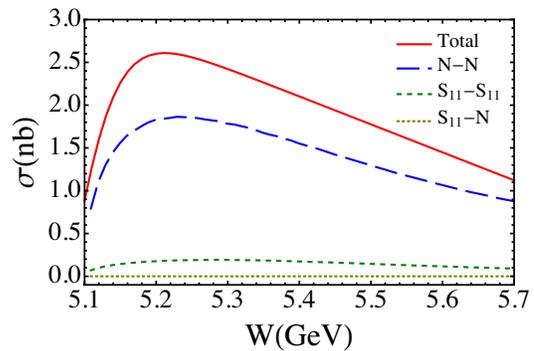}
\caption{(Color online) Total cross section for $d+d\to K^+ + K^+ +(n,n,\Lambda,\Lambda$). From upper to lower the curves stand for the total cross sections of the full calculations, exclusive process from the nucleon Born terms, exclusive process from the double $S_{11}(1535)$ excitations, and exclusive process from the one Born transition and one $S_{11}(1535)$ excitation.}
\label{fig:cross}
\end{figure}

Taking into account the large $\pi NN$ coupling and the small pion mass, the underlying mechanism will be dominated by the $\pi^0$ exchange and through two combined elementary process $\pi^0+p\to\Lambda + K^+$, which is known. The Born term, with a virtual nucleon in the intermediate state, gives a significant contribution, as seen in Fig.~\ref{fig:cross}. The $S_{11}(1535)$ resonance has relatively smaller contributions mainly due to the smaller couplings to $\pi N$ and $K\Lambda$.   Although some higher $N^*$ resonances may also have important contributions  and benefit from a significant coupling to $K\Lambda$, our estimate including the Born term and $S_{11}(1535)$ excitations can be regarded as a conservative estimate of the $T$ production cross section.

\begin{figure}[!htbp]
\centering
\includegraphics[width=0.8\linewidth]{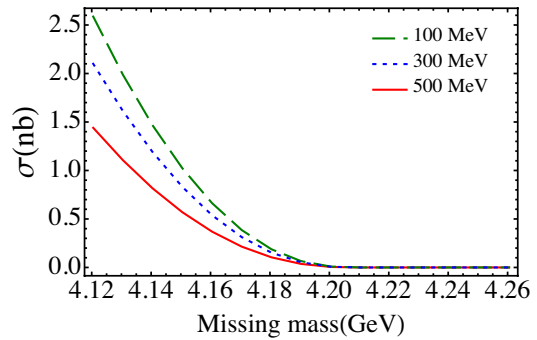}
\caption{(Color online) The missing mass spectra for the recoiled $(n,n,\Lambda,\Lambda)$ at different energies above the production threshold in  $d+d\to K^+ + K^+ +(n,n,\Lambda,\Lambda)$}.
\label{fig:missingmass}
\end{figure}

In Fig.~\ref{fig:missingmass} we show the missing mass spectra for the recoiled $(n,n,\Lambda,\Lambda)$ at different energies above the production threshold. It shows that the peak position is located at the four baryon $nn\Lambda\Lambda$ threshold. Since we have not introduced a dynamic wavefunction for  $(n,n,\Lambda,\Lambda)$, the integration over the internal momentum only present a momentum distribution that we introduced for these two $n\Lambda$ clusters. This leads to the relatively extended tail of the peak as shown in Fig.~\ref{fig:missingmass}. However, notice that the deuteron is dominated by the $S$-wave and $T$ is also treated as an $S$-wave isoscalar system. Our estimate is sufficient to demonstrate the behavior of the correlated system recoiled by the $K^+$ pair. In case that the final-state baryons are not bound, the peak will disappear in the missing mass spectrum.

\section{Summary}\label{summ}

The stability of $\isotope[3][\Lambda]{n}$ and, more likely, of $T=\isotope[4][\Lambda\Lambda]{n}$ is within the uncertainties of our knowledge of the baryon-baryon interaction. Many effects should be taken into account to refine the predictions. We already mentioned the 3-body and 4-body forces. Also, it would be interesting to unfold the effective $\Lambda N$ interaction, to separate the contribution from $\Lambda N\leftrightarrow \Sigma N\leftrightarrow \Lambda N$, and to recalculate the $(n,p,\Lambda)$ systems with explicit coupling to $(N,N,\Sigma)$. Similar considerations hold for the $\Lambda\Lambda$ coupling to $\Xi N$.    As stressed in the literature \cite{Garcilazo:2012qv,Gibson:1977ch,Hiyama:2014cua}, the coupling to channels with $\Sigma$ or $\Xi$ gives extra attraction if it induces additional spin and isospin coupling that are not present if the picture is restricted to nucleons and $\Lambda$.

As in earlier studies, e.g., \cite{Filikhin:2002wp,Nemura:2002hv}, we used simple potentials that mimic more elaborate interactions. For three-nucleons systems, this is a notorious source of overbinding, as analyzed in the literature \cite{Blatt:628052}, in particular for the tensor forces absorbed into an effective central component. However, the effect becomes much less important for dilute systems whose wave function extends very far and thus does not probe the details of the interaction region.

Most interesting, in our opinion, is the possibility to produce $T$ in deuteron-deuteron $(dd)$ collisions. As both $d$ and $T$ are weakly bound, their constituents are almost on shell. This simplifies the calculation of the production cross-section  coupling from elementary coupling among hadrons, and suppresses any off-shell corrections. We are confident that this will be an efficient doorway to this system. In brief, this new type of element, if does exist, would bring novel insights into the nuclear forces that hold the building block hadrons together and provide a great opportunity for extending our knowledge to some unreached part in our matter world.

\section*{Acknowledgement}
Useful discussions with Emiko Hiyama, Avragam Gal, Andreas Nogga  and Alfredo Valcarce are acknowledged.
This collaboration was made possible by the China--France program FCPPL. This work is also supported, in part,
by the National Natural Science Foundation of China (Grant No.
11035006), the Chinese Academy of Sciences
(KJCX3-SYW-N2), and the Sino-German CRC 110 ``Symmetries and
the Emergence of Structure in QCD" (NSFC Grant No. 11261130311).

\end{document}